\documentclass[superscriptaddress,showpacs,aps,twocolumn,prb,floatfix]{revtex4}
\usepackage{bm,amsmath,amssymb}
\usepackage{graphicx}
\usepackage{color}
\usepackage{multirow}

\begin{document}

\title{Exact analytical solution of a time-reversal-invariant topological superconducting wire}

\author{Armando A. Aligia}
\affiliation{Centro At\'omico Bariloche and Instituto Balseiro, CNEA, 8400 S. C. de Bariloche, Argentina}

\author{Alberto Camjayi}
\affiliation{Departamento de F\'{\i}sica, FCEyN, Universidad de Buenos Aires and IFIBA, 
Pabell\'on I, Ciudad Universitaria, 1428 CABA Argentina} 


\begin{abstract}
We consider a model proposed before for a time-reversal-invariant topological superconductor 
(TRITOPS) which contains a hopping term $t$, a chemical potential $\mu$, 
an extended $s$-wave pairing $\Delta$ and spin-orbit coupling $\lambda$.
We show that for $|\Delta|=|\lambda|$, $\mu=t=0$, the model has an exact analytical solution 
defining new fermion operators involving nearest-neighbor sites.
The many-body ground state is four-fold degenerate due to the existence
of two zero-energy modes localized exactly at the first and the last site 
of the chain. These four states show entanglement in the sense that
creating or annihilating a zero-energy mode at the first site is proportional 
to a similar operation at the last site. By continuity, this property should persist 
for general parameters. 
Using these results we discuss some statements related with the so called ``time-reversal anomaly''. 
Addition of a small hopping term $t$ for a chain with an even number of sites 
breaks the degeneracy and the ground state becomes unique with
an even number of particles. We also consider a small magnetic field $B$ applied to one end 
of the chain. We compare the many-body excitation energies 
and spin projection along the spin-orbit 
direction for both ends of the chains obtained treating $t$ and $B$ as small perturabtions 
with numerical results in a short chain obtaining good agreement.
\end{abstract}

\maketitle

\section{Introduction}

\label{intro}

In recent years, there have been a lot of interest in topological superconductors.
One of the main reasons 
for this attention is the existence of
Majorana fermions at the ends of one dimensional wires, 
which might be used in quantum computation 
exploiting their non-abelian nature.\cite{kitaev0,kitaev-qc}

The first theoretical proposals \cite{kitaev,wires1,wires2,fu,nadj} and experimental 
research \cite{wires-exp1,wires-exp2,wires-exp3,wires-exp4,wires-exp5} were focused on systems 
in which time-reversal symmetry is broken. More recently theoretical research
on time-reversal-invariant topological superconductors (TRITOPS) has 
developed.\cite{qi1,qi2,tritops-ort,tritops-bt,chung,fanz,kesel,dumi1,haim,je1,yaco,mat,je2,yuval,
tritops-ber,mellars,gong,cam,klino,par,entangle,sch-fu,jorg,hu,mash,tri-os,andreev,review,schrade}

The TRITOPS belong to class DIII in the classification of topological superconductors.\cite{schny}
As such, they host a zero-energy fermionic excitation,
(or equivalently a Kramers pair of Majorana fermions) at each end of the wire.
This ``Majorana Kramers Qubit'' has been proposed as the basis of a universal gate set
for quantum computing.\cite{schrade} 

Zhang \textit{et al.} \cite{fanz} proposed to construct TRITOPS wires via the proximity effect
between nodeless extended {\textit s}-wave iron-based superconductors and
semiconducting systems with large Rashba spin-orbit coupling. 
An extended {\textit s}-wave superconducting gap $\Delta$ and spin-orbit coupling $\lambda$
are the basic ingredients in the TRITOPS physics.

A remarkable property of the TRITOPS wires is that the many-body ground state is characterized by a 
fractional spin projection along the spin-orbit coupling (which we choose to be $z$) at each end of the wire.
Specifically for the left and right ends $S_\text{left}^{z},S_\text{right}^{z}= \pm 1/4$. 

The first argument to show this fractional spin \cite{kesel,review} was based on the so called
``time-reversal anomaly,'' first proposed for two- and three-dimensional TRITOPS.\cite{qi1} 
The argument can be summarized as follows. 
Denote as 
$a_{\text{left} \uparrow}$ the annihilation operator 
of the zero-energy mode at the left end of the chain. It commutes with the Hamiltonian $H$ 
and 
$[a_{\text{left} \uparrow},S_z]=(1/2)a_{\text{left} \uparrow}$, where $S_z$ is the total spin projection.
In addition, as shown below, the time reversal of 
$a_{\text{left} \uparrow}$
is proportional to 
$a_{\text{left} \uparrow}^\dagger$:
\begin{equation}
K a_{\text{left} \uparrow} K^{\dagger}=a_{\text{left} \downarrow}\propto a_{\text{left} \uparrow}^\dagger
\end{equation}
with $K$ the time reversal operator, so there is only one
independent fermion at the left end, and the same happens at the right end.
Let us assume that $|G_{0}\rangle $ is one of the degenerate ground states,
with 
$a _{\text{left}\uparrow }|G_{0}\rangle =0$. Then, since 
$[a_{\text{left}\uparrow }^{\dagger },H]=0$, also 
$|G_{1}\rangle =a _{\text{left}\uparrow }^{\dagger }|G_{0}\rangle $ belongs to the ground-state manifold
(if it does not vanish). Due to time reversal invariance of the Hamiltonian,
one might expect that $|G_{0}\rangle $ and $|G_{1}\rangle $ are time reversal
partners, and this implies $\langle G_{1}|S_{\text{left}}^{z}|G_{1}\rangle
=-\langle G_{0}|S_{\text{left}}^{z}|G_{0}\rangle $, 
where $S_\text{left}^{z}$ is the spin projection at the left end of the chain.\cite{entangle}
On the other hand, since 
$a_{\text{left} \uparrow}$ annihilates a spin up (or creates a spin down)
at the left of the chain $\langle G_1| S_\text{left}^{z} |G_1 \rangle - \langle G_0| S_\text{left}^{z} |G_0 \rangle= -1/2$,
and then $\langle G_0| S_\text{left}^{z} |G_0 \rangle=1/4$. In this argument, neither the right end 
nor the whole degeneracy of the ground state is considered.
Our results, in which the many-body states are constructed explicitly, shed light on 
the underlying physics (see Section \ref{tra}). 
We obtain that $K|G_{0}\rangle $ is not proportional to $|G_{1}\rangle$ but coincides with a third ground state.

In a previous publication, the excitations at both ends 
were studied.\cite{entangle} The contribution of each site
to the operators 
$a_{\text{left} \sigma}$ and $a_{\text{right} \sigma}$
 decays exponentially with the distance to the corresponding end, with
a decay length $\lambda_e$ determined by solving a quartic equation. For the particular 
case of the chemical potential $\mu=0$ and the length of the chain $L \rightarrow \infty$,
an analytical form of the end operators was given. Formally, one of the many-body states $|e_1 \rangle$ 
that is part of the ground-state manifold for $L \rightarrow \infty$ 
is constructed in a similar way as the ground state of 
the Bardeen-Cooper-Schrieffer (BCS) Hamiltonian, as the product of all annihilation operators 
$\Gamma _{\nu }$
satisfying $[\Gamma _{\nu },H]=E_{\nu }\Gamma _{\nu }$ with positive $E_{\nu }$. 
For finite $L$, an exponentially small mixing of
$a_{\text{left} \sigma}$ and $a_{\text{right} \sigma}$
takes place. Two new mixed annihilation operators $\gamma_{\sigma}$ with $ [\gamma_{\sigma},H]=E_{\sigma } \gamma_{\sigma}$ 
with positive $E_{\sigma }$ are found, so that the (now non-degenerate) ground state becomes
$|g_e \rangle = \gamma_{\uparrow} \gamma_{\downarrow} |e_1 \rangle$. 
Although the explicit form of $|e_1 \rangle$ is not known, the facts that it is time reversal invariant 
in the absence of a magnetic field and that the operators $\Gamma _{\nu }$ correspond to finite energy
have been used to calculate the spin projection $S_\text{right}^{z}$ for the ground state
and the first excited states with odd number of particles, in particular for a magnetic field applied only 
to the right end, finding fractional values.

Some open questions regarding the nature of the many-body ground state still remain. For example, for 
$L \rightarrow \infty$ there are two independent zero-energy modes, one at 
each end of the chain.
Then, one expects a four-fold degenerate ground state depending on
whether the occupation number of these two fermions is zero or one. However, in principle there are 
16 
combinations of  
the operators $\Gamma_{\text{left} \sigma}$ and 
$\Gamma_{\text{right} \sigma}$ that could be applied to $|e_1 \rangle$. How are they related? 
To answer this question the explicit form of 
$|e_1 \rangle$ is needed. 

In this work, we report on the exact analytical solution of the model for particular parameters ($|\Delta|-|\lambda|=\mu=t=0$), 
with $L$ finite and open boundary conditions.
This allows us to construct explicitly, not only the one-body operators that diagonalize the Hamiltonian, but also 
the four many-body states that are part of the ground state.
We find that $\Gamma_{\text{right} \sigma} |e_1 \rangle$ is proportional to $\Gamma_{\text{left} \sigma} |e_1 \rangle$, where 
$|e_1 \rangle$ is constructed as indicated above. By continuity, this property should be valid for general
parameters inside the topological phase ($|\mu| < 2 |\lambda|$). This indicates that although  
$|e_1 \rangle$ does not seem to contain information about the zero-energy modes (it is constructed 
with operators that commute with the zero-energy ones), it is in fact an entangled state and its ends
are related. 

Using perturbative methods, 
we also discuss the effect of a small hopping $t$ and a magnetic field applied to one end on 
the exact solution. The analytical results for the many-body states are supported by 
numerical diagonalization of small systems.

The paper is organized as follows. In Sec. \ref{model} the model is
described. In Sec. \ref{exact} we construct the exact  analytical solution of the model for
$|\Delta|-|\lambda|=\mu=t=0$ and describe the many-body ground state. 
We also calculate the expectation values of the spin projection at the ends
for the different states that compose the ground state, or appropriate linear combinations of them,  
finding the fractional values $\pm 1/4$ 
expected from previous works. In Sec. \ref{hop} we analyze perturbatively the effect
of a small hopping $t$ on the exact solution and compare with numerical results. 
In Sec. \ref{sum} we present a summary and a brief discussion.

\section{Model}

\label{model}

The Hamiltonian 
describing the system reads\cite{fanz,cam}

\begin{multline}
H = \sum_{\sigma }\left\{ \sum_{j=1}^{L-1} \left(-tc_{j+1\sigma }^{\dagger }c_{j\sigma}
+is_{\sigma }\lambda c_{j+1\sigma }^{\dagger }c_{j\sigma } \right. \right.  \\
\left. \left. + s_{\sigma}\Delta e^{i\phi }c_{j+1\sigma }^{\dagger }c_{j\overline{\sigma }}^{\dagger
}+\mathrm{H.c.}\right)-\mu \sum_{j=1}^{L} c_{j\sigma }^{\dagger }c_{j\sigma } \right\},  \label{ham}
\end{multline}
where $s_{\uparrow ,\downarrow }=\pm 1$ and $\overline{\uparrow }=\downarrow
,\;\overline{\downarrow }=\uparrow $. The parameter $t$ corresponds to the
nearest-neighbor hopping, $\mu $ is the chemical potential, $\lambda $ is
the Rashba spin-orbit coupling and $\Delta $ is the strength of the extended
{\textit s}-wave pairing.

In the
following we will take for simplicity $\Delta =\lambda =1$, $\phi =t=\mu =0$. The
formalism used can be changed in a straightforward way to include a finite
phase $\phi $ and other signs of $\Delta $ and $\lambda $. The effect of
finite $t$ is treated in Section \ref{hop}, and the general case is discussed in
Section \ref{sum}.

For $\phi =0$, the Hamiltonian is invariant under time reversal symmetry. In
addition, the Hamiltonian conserves parity and the total spin projection $S_{z}$ 
(the total spin in the direction of the Rashba spin-orbit coupling).
Below we will use these three symmetries.

\section{Construction of the exact analytical solution}

\label{exact}

\subsection{One-particle operators}

\label{exact1}

In this Section, we look for annihilation operators $\Gamma _{\nu }$ 
satisfying

\begin{equation}
\lbrack \Gamma _{\nu },H \rbrack = E_{\nu }\Gamma _{\nu },  \label{conmu0}
\end{equation}%
that diagonalize the Hamiltonian Eq.~(\ref{ham}) for $%
\Delta =\lambda =1$, $\phi =t=\mu =0$. Since Eq.~(\ref{conmu0}) implies that $\Gamma _{\nu }^{\dagger}$ 
satisfies the same equation with the opposite sign of $E_{\nu }$, we can redefine the operators so that
$E_{\nu } \ge 0$, permuting $\Gamma _{\nu }$ and $\Gamma _{\nu }^{\dagger}$ if necessary.

We define the following operators
\begin{eqnarray}
a_{j\sigma } &=&(c_{j\sigma }+ic_{j\overline{\sigma }}^{\dagger })/\sqrt{2},
\notag \\
b_{j\sigma } &=&(c_{j\sigma }-ic_{j\overline{\sigma }}^{\dagger })/\sqrt{2},
\label{sd}
\end{eqnarray}%
with the property 

\begin{equation}
a_{j\sigma }^{\dagger }=-ia_{j\overline{\sigma }},\text{ }b_{j\sigma
}^{\dagger }=ib_{j\overline{\sigma }}.  \label{conj}
\end{equation}%

Using Eq.~(\ref{ham}), the following commutators can be obtained
for a finite chain of $L$ sites:

\begin{eqnarray}
\left[ a_{j\sigma },H\right] &=&
 \begin{cases}
   2s_{\sigma }i b_{j-1\sigma } & \text{if } j>1 \\
   0                            & \text{otherwise }
 \end{cases} \notag\\
\left[ b_{j\sigma },H \right] &=&
 \begin{cases}
  -2s_{\sigma }ia_{j+1\sigma }  & \text{if } j<L \\ 
  0                             & \text{otherwise}  \label{conmu1}
\end{cases}
\end{eqnarray}%
with $j=1,\ldots,L$

\begin{figure}[h]
\includegraphics[width=\columnwidth]{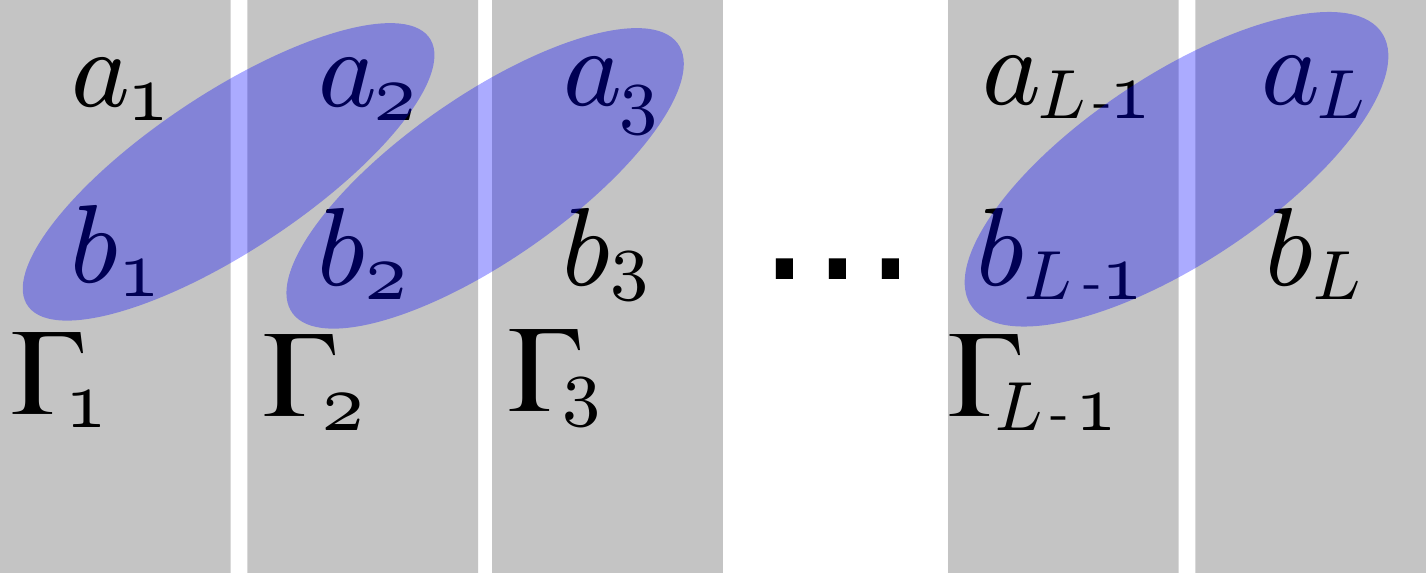}
\caption{(Color online) Sketch of the construction of the annihilation operators of finite energy,
leaving the zero-energy modes $a_{1 \sigma}$ and $b_{L \sigma}$ at the ends.}
\label{fig}
\end{figure}

Clearly $a_{1\sigma }$ and $b_{L\sigma }$, taking into account Eqs.~(\ref{conj}), 
correspond to the zero-energy modes, one for each end of the chain,
expected from the topological character of the TRITOPS phase.\cite{fanz,cam} 
The remaining operators mix nearest-neighbor sites as sketched in Fig.~\ref{fig},
resembling the exact  analytical solution of the Kitaev model.\cite{kitaev}

Defining for $j<L$ the operators

\begin{eqnarray}
\Gamma _{j \sigma } &=&(b_{j \sigma }-i s_{\sigma} a_{j+1 \sigma })/\sqrt{2},  \label{gamma}
\end{eqnarray}
it is easy to see, using Eqs.~(\ref{conmu1}), that their commutators with the Hamiltonian are

\begin{equation}
\lbrack \Gamma _{j\sigma },H]=2\Gamma _{j\sigma }.  \label{conmu2}
\end{equation}%
Therefore the desired annihilation operators are obtained.

Note that, under time reversal $K$, the operators defined so far transform as

\begin{equation}
K\alpha _{j\uparrow }K^{\dagger }=\alpha _{j\downarrow },\text{ }K\alpha
_{j\downarrow }K^{\dagger }=-\alpha _{j\uparrow },  \label{k}
\end{equation}
where $\alpha =a$, $b$ or $\Gamma$.

\subsection{Many-body low-energy eigenstates}

Since there are two independent fermionic modes with zero energy, one at
each end of the chain, one expects a four-fold degenerate ground state
depending on the occupancy of these modes. The analysis 
below as
well as many-body calculations in chains with up to $L=6$ sites confirm this
expectation.

One of these four states, with even number of particles is obtained applying
all annihilation operators with positive energy [those entering 
Eq.~(\ref{conmu2})] to the vacuum $|0\rangle $ of the $c_{j\sigma }$:

\begin{equation}
|e_1\rangle =N_{L}\prod\limits_{j=1}^{L-1}\Gamma _{j\uparrow }\Gamma
_{j\downarrow }|0\rangle ,  \label{e1}
\end{equation}%
where $N_{L}$ is a normalization factor. This state is invariant under time
reversal [using Eqs.~(\ref{k}) one easily proves that $K|e_1\rangle
=|e_1\rangle $] and also $S_{z}|e_1\rangle =0$, where $S_{z}=\sum_{j=1}^{L} S_{j}^{z}$ is the total spin projection. 
Note that neither the Hamiltonian nor this state have inversion symmetry.

Two states with odd parity number can be written as
\begin{equation}
|o\sigma \rangle =N_{o}a_{1\overline{\sigma }}|e_1\rangle ,  \label{odd}
\end{equation}%
where the normalization factor $N_{o}=\sqrt{2}$ as shown below Eq. (\ref{szpunta}). 
These states are also
eigenstates of the total spin projection with $S_{z}|o\sigma \rangle
=(s_{\sigma }/2)|o\sigma \rangle $. 

Finally, there is another even state
invariant under time reversal and with zero spin projection,

\begin{equation}
|e_2\rangle =\left( a_{1\uparrow }a_{1\downarrow }-a_{1\downarrow
}a_{1\uparrow }\right) |e_1\rangle .  \label{e2}
\end{equation}%
Using Eqs.~(\ref{conj}) it is easy to see that this state is normalized ($\langle e_2|e_2\rangle=1$) 
and orthogonal to $|e_1\rangle $. In fact,
except for normalization factors, $\langle e_1|e_2\rangle \sim
i(\langle o\uparrow |o\uparrow \rangle -\langle o\downarrow |o\downarrow
\rangle )=0$.
Therefore, all these states constitute an orthonormal basis of the ground state manifold.

It might seem surprising that acting with the zero-mode operators at the right end, 
$b_{L\sigma }$ on $|e_1\rangle $, does not lead to new states that are part of
the ground state. Instead, we find
\begin{eqnarray}
b_{L\sigma }|e_1\rangle &=& -(s_{\sigma} i)^{L-1} a_{1 \sigma}|e_1\rangle, \notag \\
a_{1 \sigma}|e_1\rangle &=& -(s_{\overline{\sigma}} i)^{L-1}b_{L\sigma }|e_1\rangle. \label{puntas}
\end{eqnarray}%

This can be proved by induction. 
For the simplest chain with $L=2$, the ground state manifold is
\begin{eqnarray}
|e_1\rangle &=&\frac{1}{2}[c_{1\uparrow }^{\dagger }c_{1\downarrow }^{\dagger
}+c_{2\uparrow }^{\dagger }c_{2\downarrow }^{\dagger }+ic_{1\uparrow
}^{\dagger }c_{2\downarrow }^{\dagger }+ic_{1\downarrow }^{\dagger
}c_{2\uparrow }^{\dagger }]|0\rangle ,  \notag \\
|e_2\rangle &=&-\frac{1}{2}[1-c_{1\uparrow }^{\dagger }c_{2\downarrow
}^{\dagger }+c_{1\downarrow }^{\dagger }c_{2\uparrow }^{\dagger
}-c_{1\uparrow }^{\dagger }c_{1\downarrow }^{\dagger }c_{2\uparrow
}^{\dagger }c_{2\downarrow }^{\dagger }]|0\rangle ,  \notag \\
|o \downarrow \rangle &=&\frac{1}{2}[c_{1\downarrow }^{\dagger
}+ic_{2\downarrow }^{\dagger }+ic_{1\downarrow }^{\dagger }c_{2\uparrow
}^{\dagger }c_{2\downarrow }^{\dagger }+c_{1\uparrow }^{\dagger
}c_{1\downarrow }^{\dagger }c_{2\downarrow }^{\dagger }]|0\rangle ,  \notag
\\
|o \uparrow \rangle &=& K|o\downarrow \rangle.  \label{l2}
\end{eqnarray}%
Using these expressions, it is easy to check that Eq.~(\ref{puntas}) is
valid for $L=2$. 

Now we prove its validity for a chain of $L+1$ sites
assuming that Eq.~(\ref{puntas}) 
holds for $L$ sites.

For a chain of $L+1$ sites, Eq.~(\ref{e1})
can be written in the form

\begin{eqnarray}
|e_1(L+1)\rangle &=&N_{L+1}\hat{O}(L+1)|0\rangle ,  \notag \\
\hat{O}(L) &=&\prod\limits_{j=1}^{L-1}\Gamma _{j\uparrow }\Gamma _{j\downarrow }.
\label{elp1}
\end{eqnarray}
Then using anticommutation rules and Eq.~(\ref{puntas})%
\begin{multline}
a_{1\uparrow }|e_1(L+1)\rangle = N_{L+1}\Gamma _{L\uparrow }\Gamma
_{L\downarrow }a_{1\uparrow }\hat{O}(L)|0\rangle \\
= -(-i)^{L-1}N_{L+1}\Gamma _{L\uparrow }\Gamma _{L\downarrow
}b_{L\uparrow }\hat{O}(L)|0\rangle \\
=(-i)^{L-1}N_{L+1}\hat{O}(L)\Gamma _{L\downarrow }\Gamma _{L\uparrow
}b_{L\uparrow }|0\rangle .  \label{auxi}
\end{multline}%
From the definitions Eqs.~(\ref{sd}) and (\ref{gamma}) we find

\begin{equation}
\Gamma _{L\uparrow }b_{L\uparrow }|0\rangle =ib_{L+1\uparrow }\Gamma
_{L\uparrow }|0\rangle =\frac{-i}{2\sqrt{2}}c_{L+1\downarrow }^{\dagger
}c_{L\downarrow }^{\dagger }|0\rangle ,  \label{auxi2}
\end{equation}%
and replacing it in Eq.~(\ref{auxi})

\begin{multline}
a_{1\uparrow }|e_1(L+1)\rangle =-(-i)^{L}N_{L+1}\hat{O}(L)\Gamma_{L\downarrow
}b_{L+1\uparrow }\Gamma _{L\uparrow }|0\rangle \\
 = -(-i)^{L}N_{L+1}b_{L+1\uparrow }\Gamma _{L\uparrow }\Gamma
_{L\downarrow }\hat{O}(L)|0\rangle  \\
= -(-i)^{L}b_{L+1\uparrow }|e_1(L+1)\rangle ,  \label{puntas2}
\end{multline}%
in agreement with Eq.~(\ref{puntas}). The corresponding relation for the
opposite spin of the end operators is obtained using the time-reversal
operator $K$.

\subsubsection{Ground-state energy}

The energy of the four-fold degenerate ground state $E_{g}$ can be obtained
from the following argument. Let us define the charge conjugation (or
electron-hole transformation) $C$ as the one which permutes annihilation and
creation operators

\begin{equation}
Cc_{j\sigma }^{\dagger }C=c_{j\sigma },\text{ }Cc_{j\sigma }C=c_{j\sigma
}^{\dagger }.  \label{chc}
\end{equation}

Clearly $C^{2}=1.$ For $\phi =0$, the Hamiltonian Eq.~(\ref{ham}) transforms
as

\begin{equation}
CHC=-2\mu L-H.  \label{ch}
\end{equation}%
In the case we are considering, with $\mu =0$, this implies that if a many-body state $|i\rangle 
$ is an eigenstate with energy $E_{i}$, its electron-hole partner $%
C|i\rangle $ is also an eigenstate with energy $-E_{i}$.
%
Thus, the spectrum is symmetric around zero energy.

Since the system is non interacting, the excited states are obtained
applying creation operators $\Gamma _{j\uparrow }^{\dagger }$ (each with an
energy cost 2) and zero-energy operators (without energy cost) to $|e_1\rangle $. 
Clearly one of the states of highest energy $E_\mathrm{max}$ is 
$\prod\limits_{j=1}^{L-1}\Gamma _{j\uparrow }^{\dagger }\Gamma _{j\downarrow
}^{\dagger }|e_1\rangle $, and $E_\mathrm{max}-E_{g}=4(L-1)$. In addition, since the
total spectrum is symmetric $E_\mathrm{max}+E_{g}=0$. Thus, the ground-sate energy
is
\begin{equation}
E_{g}=-2(L-1).  \label{eg}
\end{equation}%
This has been confirmed by numerical calculations in small systems.

\subsubsection{Expectation value of the spin projection at both ends}
\label{szfrac}

Proceeding in a way similar as shown 
above, it is possible to 
write the state $|o\downarrow \rangle $ (except for a phase) as

\begin{multline}
b_{L\uparrow }|e_1(L)\rangle =-N_{L}\hat{O}(L-1)\Gamma _{L-1\downarrow
}b_{L\uparrow }\Gamma _{L-1\uparrow }|0\rangle \\
= \frac{-1}{4\sqrt{2}}N_{L}\hat{O}(L-1)\left[-c_{L\downarrow }^{\dagger
}+ic_{L-1\downarrow }^{\dagger }+\right. \\
\left.+ic_{L-1\uparrow }^{\dagger }c_{L-1\downarrow }^{\dagger }c_{L\downarrow}^{\dagger }
+c_{L-1\downarrow }^{\dagger }c_{L\uparrow
}^{\dagger }c_{L\downarrow }^{\dagger }\right]|0\rangle .  \label{bl2}
\end{multline}%
From here it is easy to see that the expectation value of the spin
projection at the right end [for the parameters of the exact  analytical solution, 
$S_\text{right}^{z}= S_{L}^{z}=(c_{L\uparrow }^{\dagger
}c_{L\uparrow }-c_{L\downarrow }^{\dagger }c_{L\downarrow
})/2$] becomes

\begin{equation}
\langle o\downarrow |S_\text{right}^{z}|o\downarrow \rangle =-1/4,  \label{szpunta}
\end{equation}%
since half of the terms of Eq.~(\ref{bl2}) contribute with $-1/2$ and the
other half do not contribute. 
In addition, comparing the norm of $b_{L\uparrow }\Gamma _{L-1\uparrow }\Gamma _{L-1\downarrow }|0\rangle $ and 
$\Gamma _{L-1\uparrow }\Gamma _{L-1\downarrow }|0\rangle $, and using Eq. (\ref{puntas}) 
one realizes that the norm entering Eq. (\ref{odd}) is $N_{o}=\sqrt{2}$.

Similarly, with $S_\text{left}^{z}= S_{1}^{z}$,
\begin{eqnarray}
\langle o\downarrow|S_\text{left}^{z}|o\downarrow \rangle &=& 
\langle o\downarrow|S_\text{right}^{z}|o\downarrow \rangle= -1/4,  \notag \\
\langle o\uparrow|S_\text{left}^{z}|o\uparrow \rangle &=&
\langle o\uparrow|S_\text{right}^{z}|o\uparrow \rangle =1/4,  \label{szp2}
\end{eqnarray}%
in agreement with previous results derived for the general case.\cite{entangle}
These results might be expected from Eq.~(\ref{puntas}). In fact $S_{z}|e_1 \rangle=0$
and $S_{z}|o\downarrow \rangle=S_{z}\sqrt{2}a_{1\uparrow }|e_1 \rangle=-\frac{1}{2}|o\downarrow \rangle$,
as can be easily shown
using $[S_{z},a_{1\uparrow }]=-a_{1\uparrow }/2$. Then $a_{1\uparrow }$ changes the total spin 
projection by $-1/2$ and, since it is a local operator, one would naively expect that this change 
is concentrated at the left of the chain. However, since the application of a zero mode operator 
with the same spin projection at the left or at the right end of the chain have the same physical effect, 
the change in the total spin is actually equally shared by both ends.
A similar argument can be applied for $|o\uparrow \rangle$.

Concerning the expectation values of $S_\text{left}^{z}$ and $S_\text{right}^{z}$ 
for the states $|e_1\rangle $ and $|e_2\rangle $, they are zero since
both states are
time-reversal invariant. However, this is no 
longer true for an 
arbitrary linear
combination. Using $K|e_{i}\rangle =|e_{i}\rangle $, $i=1,2$; 
$KS_{j}^{z}K^{\dagger }=-S_{j}^{z}$, $j=1,\dots,L$, and the antiunitary
properties of $K$ we have 

\begin{eqnarray}
\langle e_{i}|S_{j}^{z}|e_{k}\rangle &=&\langle e_{i}|(K^{\dagger}KS_{j}^{z}K^{\dagger }K|e_{k}\rangle  \notag \\
&=& \overline{\langle e_{i}|K^{\dagger })(KS_{j}^{z}K^{\dagger}K|e_{k}\rangle}=-\overline{\langle e_{i}|S_{j}^{z}|e_{k}\rangle }  \notag
\\
&=&-\langle e_{k}|S_{j}^{z}|e_{i}\rangle ,  \label{szme}
\end{eqnarray}
where the parenthesis separate the operators acting to the right and to the left of the parenthesis. 
Clearly 
$\langle e_{i}|S_{j}^{z}|e_{i}\rangle =0$ and $\langle e_{k}|S_{j}^{z}|e_{i}\rangle $ 
for $i \neq k$ is a pure imaginary number. It is easy to see
that the absolute value of the expectation value of the spin at any site (in
particular at the ends $j=1$, $L$) is maximized by the following linear
combinations

\begin{equation}
|e_{3}\rangle =\frac{1}{\sqrt{2}}\left( |e_{1}\rangle +i|e_{2}\rangle
\right) ,\text{ }|e_{4}\rangle =\frac{1}{\sqrt{2}}\left( |e_{1}\rangle
-i|e_{2}\rangle \right) ,  \label{e34}
\end{equation}
with

\begin{equation}
\langle e_{_{3(4)}}|S_{j}^{z}|e_{_{3(4)}}\rangle =\pm i\langle
e_{1}|S_{j}^{z}|e_{2}\rangle .  \label{sz34}
\end{equation}

Using Eqs.~(\ref{conj}), (\ref{e2}), (\ref{odd}), (\ref{szp2}), and 
$[S_{1}^{z},a_{1\uparrow }^{\dagger }]=a_{1\uparrow }^{\dagger }/2$,

\begin{eqnarray}
\langle e_{1}|S_{1}^{z}|e_{2}\rangle &=&\langle e_{1}|S_{1}^{z}\left(
a_{1\uparrow }a_{1\downarrow }-a_{1\downarrow }a_{1\uparrow }\right)
|e_{1}\rangle  \notag \\
&=&i\langle e_{1}|S_{1}^{z}\left( 1-2a_{1\uparrow }^{\dagger }a_{1\uparrow
}\right) |e_{1}\rangle  \notag \\
&=&-i\langle e_{1}|a_{1\uparrow }^{\dagger }(2S_{1}^{z}+1)a_{1\uparrow
}|e_{1}\rangle =  \notag \\
&=&\frac{-i}{2}\langle o\downarrow |(2S_{1}^{z}+1)|o\downarrow \rangle =%
\frac{-i}{4}.  \label{szmix}
\end{eqnarray}
This result is easily verified for $L=2$ using Eqs.~(\ref{l2}).

Replacing Eq.~(\ref{szmix}) in Eq.~(\ref{sz34}) and using $\langle
e_{_{i}}|S_{j}^{z}|e_{i}\rangle =0$ for $j \neq 1 \mbox{ or } L$,
 we finally obtain

\begin{eqnarray}
\langle e_{3}|S_{\text{left}}^{z}|e_{3}\rangle  &=&-\langle e_{3}|S_{\text{right}}^{z}|e_{3}\rangle =1/4,  \notag \\
\langle e_{4}|S_{\text{left}}^{z}|e_{4}\rangle  &=&-\langle e_{4}|S_{\text{right}}^{z}|e_{4}\rangle =-1/4  \label{mix}
\end{eqnarray}

From these equations, it is clear that application of a magnetic field in the direction of the spin-orbit
coupling ($z$) applied at one of the ends only, breaks the degeneracy
of the ground state favoring one of the states $|e_{3}\rangle $, $|e_{4}\rangle$.
The state of lower energy depends on the direction of the field and the end
at which it is applied. 
This point is discussed further in Section \ref{field}. 

\subsubsection{Discussion on the ``time-reversal anomaly''}

\label{tra}

The explicit construction of the many-body eigenstates allows us to discuss
in detail the arguments involved in the so called ``time-reversal anomaly.''~\cite{qi1,kesel,review} 

As discussed in Sec.~\ref{intro}, we can identify an independent zero-mode
operator at the left end of the chain $a_{1\uparrow}=ia_{1\downarrow }^{\dagger }$, 
where we have used Eq.~(\ref{conj}) in the equality. 
For the ground state $|G_{0}\rangle $, with the property $a_{1\uparrow}|G_{0}\rangle =0$, 
$|G_{0}\rangle =|o\downarrow \rangle $ can
be chosen [see Eqs.~(\ref{e1}), (\ref{odd}), and (\ref{e2})]. 

A local fermion parity operator is defined as $P_{\text{left}}=2a_{1\uparrow}^{\dagger }a_{1\uparrow}-1$.
Clearly $P_{\text{left}}|G_{0}\rangle =-|G_{0}\rangle $. 
It has been shown
that $P_{\text{left}}$ anticommutes with the time reversal operator $K$.\cite{qi1,kesel,review} 
In the present case, this is proved using Eqs.~(\ref{conj}) and (\ref{k}).
This implies that the time reversal partner of $|G_{0}\rangle $ 
($|o\uparrow \rangle $) is even under $P_{\text{left}}$

\begin{equation}
P_{\text{left}}K|G_{0}\rangle =-KP_{\text{left}}|G_{0}\rangle
=K|G_{0}\rangle .
\end{equation}

Since time reversal should clearly commute with the total number of fermions in the system, the fact that 
$\left\{K,P_{\text{left}}\right\}=0$ 
(note that one might expect that in a semi-infinite chain, 
$P_{\text{left}}$ is the total fermion parity of the system), 
it is known as the ``time-reversal anomaly''. 
To resolve this apparent contradiction,
the other end of the chain must be considered. 
Defining the corresponding local operator at the right 
$P_{\text{right}}=-(2b_{L\uparrow}^{\dagger }b_{L\uparrow }-1)$ (as in Ref. \onlinecite{chung}), 
the product $P = P_{\text{left}}P_{\text{right}}$
coincides with the total fermion parity and, while 
$\left\{K,P_{\text{left}}\right\}=\left\{K,P_{\text{right}}\right\}=0$, $[K,P]=0$. 
This is in complete agreement with previous arguments.\cite{qi1,kesel,review} 

\begin{figure}[h]
\includegraphics[width=\columnwidth]{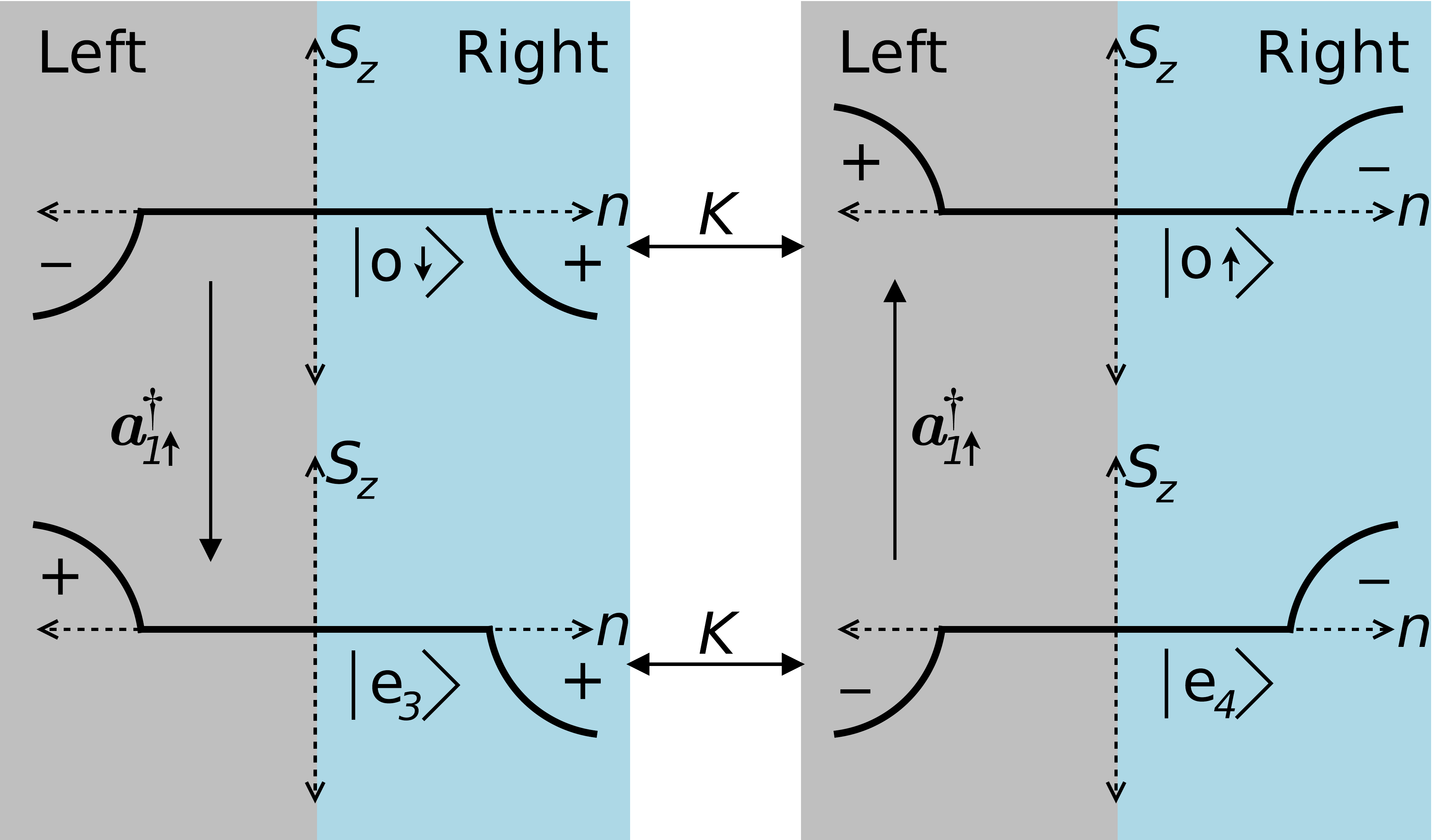}
\caption{(Color online) Many-body ground-state manifold. 
For the four orthonormal many-body ground states, we sketch the $S_z$ value at each chain site $n$. 
The local parity ($P_\text{left}$, $P_\text{right}$) value is
indicated with $+/-$ at the corresponding end. 
The states are accommodated such that left and right panels are related 
by time reversal symmetry and top and bottom states are connected by the action of $a^{\dagger}_{1\uparrow}$.}
\label{fig2}
\end{figure}

However, for systems which conserve one component of the total spin ($S_z$ in our case), 
an additional argument, related to the \textquotedblleft time-reversal anomaly \textquotedblright\, 
is usually invoked to explain the 
spin expectation value at the end of the chain.\cite{review}
Let $|G_{0}\rangle $ be one of the degenerate ground states with 
$a _{1 \uparrow }|G_{0}\rangle =0$. Then also 
$|G_{1}\rangle =a _{1\uparrow }^{\dagger }|G_{0}\rangle $ belongs to the ground state manifold. 
On the other hand, since 
$a_{1 \uparrow}$ annihilates a spin up (or creates a spin down),
at the left of the chain $\langle G_1| S_\text{left}^{z} |G_1 \rangle - \langle G_0| S_\text{left}^{z} |G_0 \rangle= -1/2$.
Due to time reversal invariance of the Hamiltonian,
one might expect that $|G_{0}\rangle $ and $|G_{1}\rangle $ are time reversal
partners, $\langle G_{1}|S_{\text{left}}^{z}|G_{1}\rangle
=-\langle G_{0}|S_{\text{left}}^{z}|G_{0}\rangle $, 
and then $\langle G_0| S_\text{left}^{z} |G_0 \rangle=1/4$.

The incorrect assumption in the argument discussed above is to associate the ground state $%
|G_{1}\rangle$ with the time-reversal partner of 
$|G_{0}\rangle $. 
In fact, as shown previously, it is possible to choose 
$|G_{0}\rangle =|o\downarrow \rangle$. Algebraic manipulations similar
to those used in Section~\ref{szfrac} lead to $|G_{1}\rangle =|e_{3}\rangle$,
given by the first Eq.~(\ref{e34}). This is because the
ground state is four-fold degenerate and contains more states than just $|G_{0}\rangle $ 
and $K|G_{0}\rangle = |o\uparrow \rangle $.

Other possible choice of $|G_{0}\rangle $ (see Fig.~\ref{fig2}),
with the property $a_{1\uparrow}|G_{0}\rangle =0$, is 
$|e_{4}\rangle $ given by the second Eq.~(\ref{e34}). 
In this
case $K|G_{0}\rangle =|e_{3}\rangle $, which is also orthogonal to 
$|G_{1}\rangle =a_{1\uparrow}^{\dagger }|G_{0}\rangle $. An important difference between
the choices $|o\downarrow \rangle $ or $|e_{4}\rangle $ is that 
$K^{2}|o\downarrow \rangle =-|o\downarrow \rangle $
($K^{2}|e_{4}\rangle =|e_{4}\rangle $) as expected for a system with an odd
(even) number of particles. Note that for both possible choices of $|G_{0}\rangle $, $|G_{0}\rangle $
and $|G_{1}\rangle $ have opposite expectation value of $S_{\text{left}}^{z}$ (see Fig.~\ref{fig2})
but the same value of $S_{\text{right}}^{z}$, as shown in Section \ref{szfrac}. 
This is a physical indication that $|G_{0}\rangle $ and $|G_{1}\rangle $ cannot be
time-reversal partners. 

Another shortcoming in the argument is that $a_{1\uparrow }^{\dagger }$, 
being a local operator, should change the expectation value of $S_{\text{left}}^{z}$ 
(not only the total $S_{z}$) by 1/2, \emph{i.e.} the support of $a_{1\uparrow }^{\dagger }$ is near the boundary. 
This is in fact true when $a_{1\uparrow }^{\dagger }$ is
applied to $|G_{0}\rangle = |o\downarrow \rangle$ or $|e_4 \rangle$ but not in general.
When applied to $|e_{1}\rangle $ for example, since rather surprisingly 
$ a_{1\uparrow}^{\dagger}|e_{1}\rangle \propto b_{L\uparrow}^{\dagger}|e_{1}\rangle \propto |o\uparrow \rangle$, 
implying that while $\langle e_{1}|S_{\text{left}}^{z}|e_{1}\rangle
=\langle e_{1}|S_{\text{right}}^{z}|e_{1}\rangle =0$, one has $\langle o\uparrow
|S_{\text{left}}^{z}|o\uparrow \rangle =\langle o\uparrow |S_{\text{right}}^{z}|o\uparrow \rangle =1/4$, so 
that the spin is actually distributed among both chain ends (see Fig.~\ref{fig2}).

The fact that the property 
$\langle G_{1}|S_{\text{left}}^{z}|G_{1}\rangle=-\langle G_{0}|S_{\text{left}}^{z}|G_{0}\rangle$
still holds true, even when they are not time reversal partners as normally argued, is the reason
why the \textquotedblleft time-reversal anomaly \textquotedblright\ is useful to guess the fractional spin at the chain boundaries.
This relation however is not due to a symmetry connecting the states but a particularity 
of the ground state manifold.      


Previous uses of $\left\{K,P_{\text{left}}\right\}=0$ include Josephson junctions
with topological superconductors.
Inserting the chain in a ring with a magnetic flux one expects a circulating
current in the system. The flux breaks the time-reversal symmetry except for
particular values, including zero.  As argued first in Ref. \onlinecite{chung} for a TRITOPS
with a Hamiltonian different as ours,  the time-reversal symmetry is
spontaneously broken at zero flux for an odd total fermion parity $P$
leading to a spontaneous circulating current, while this does not happen for
even parity [see Eqs.~(7) of Ref. \onlinecite{chung}]. While an argument similar to the
above mentioned \textquotedblleft time-reversal anomaly \textquotedblright\
has been invoked to show this fact, it can also be derived from more general
arguments, since the states with odd fermion parity have $K^{2}=-1$, and
therefore for any such state $|o\rangle $, $K|o\rangle \neq c|o\rangle $ for
any $c$ number $c$. Instead for even parity one can construct states such as 
$K|e\rangle =|e\rangle $. The ground states of our system [see Eqs. (\ref{e1}), 
(\ref{odd}), and (\ref{e2}), and Eqs. (\ref{l2})  for the particular case 
$L=2$] are clear examples. Using $K|e\rangle =|e\rangle $, since the current
operator $I$ is odd under time reversal, one can prove that $\langle e| I|e\rangle =0$ 
following a similar argument as that leading to
Eq. (\ref{szme}).

Note that the properties of the ground states under the discrete symmetries
time reversal, total fermion parity and local fermion parities, although
demonstrated for particular parameters, are valid by continuity inside the
whole topological phase for $L \rightarrow \infty$. For finite chains, in
general there is a (very small) mixing of the end zero modes that split the
ground state, as described in the next section.

\section{Effect of a small hopping term}
\label{hop}

In this section we discuss the effect of the hopping term 
$H_{t}=-t\sum_{j=1}^{L-1}\sum_{\sigma }(c_{j+1\sigma }^{\dagger }c_{j\sigma }+\mathrm{H.c.})$ on the
exact  analytical solution. To simplify the analysis we assume $t$ much smaller than
the other energy scales ($t \ll 1$).
However the
main results concerning the splitting of the degeneracy of the many-body
states are general for a long enough chain in the topological phase, as
discussed below.

\subsection{One-body operators}
\label{hop1}

For $L\geq 4$, the commutators with the hopping term of the operators defined
in Section \ref{exact1} satisfying Eq.~(\ref{conmu0}) for $t=0$, are

\begin{eqnarray}
\lbrack a_{1\uparrow },H_{t}] &=&\frac{t}{\sqrt{2}}\left( -\Gamma
_{2\uparrow }+i\Gamma _{2\downarrow }^{\dagger }\right) ,  \notag \\
\lbrack b_{L\uparrow },H_{t}] &=&\frac{t}{\sqrt{2}}\left( -i\Gamma
_{L-2\uparrow }+\Gamma _{L-2\downarrow }^{\dagger }\right) ,  \notag \\
\lbrack \Gamma _{j\uparrow },H_{t}] &=&t\Gamma _{j\downarrow }^{\dagger }+%
\frac{t}{2}\left( L_{j}+R_{j}\right),
\label{conmut}\end{eqnarray}
with
\begin{eqnarray} 
L_{j} &=&
   \begin{cases} 
      0 & \text{ if }j=1 \\
    -\sqrt{2}a_{1\uparrow } & \text{ if }j=2 \\
    -i\Gamma _{j-2\uparrow }+\Gamma _{j-2\downarrow }^{\dagger } & \text{ if }j>2
\end{cases} \\
R_{j} &=& 
  \begin{cases}
   i\Gamma _{j+2\uparrow }+\Gamma _{j+2\downarrow }^{\dagger } & \text{ if }j<L-2 \\
   i\sqrt{2}b_{L\uparrow } & \text{ if }j=L-2 \\
   0  & \text{ if }j=L-1.
\end{cases}  
\end{eqnarray}%
The corresponding relations for the operators with opposite spin are
obtained 
applying the time-reversal operator [see Eqs.~(\ref{k})].

In first-order perturbation theory, $H_{t}$ lifts the degeneracy of the $\Gamma _{j\sigma }$, 
introducing a hopping 
to next-nearest neighbors. The subspace of the operators with even $j$ remains decoupled
of the corresponding one for odd $j$.
For each subspace of operators, diagonalization of $H_{t}$ is
equivalent to solve an open chain with $M$ sites and nearest-neighbor
hopping. For $L$ odd, $M=(L-1)/2$ for both subspaces. For $L$ even, $M=L/2-1$
for the sites with even $j$
and, $M=L/2$ for the subspace of the $\Gamma _{j\sigma }$ 
with odd $j$.

By solving these equivalent problems, for arbitrary $L$, we can define
states such that

\begin{equation}
\lbrack \Gamma _{k\sigma },H]=(2+t\cos k)\Gamma _{k\sigma }.  \label{gk}
\end{equation}%
In the subspace of the operators with odd $j$ they have the form

\begin{equation}
\Gamma _{k\uparrow }^{o}=\sqrt{\frac{2}{M+1}}\sum\limits_{l=1}^{M}\sin
(kl)(i)^{l-1}\Gamma _{2l-1\uparrow },  \label{gko}
\end{equation}%
while for even $j$

\begin{equation}
\Gamma _{k\uparrow }^{e}=\sqrt{\frac{2}{M+1}}\sum\limits_{l=1}^{M}\sin
(kl)(i)^{l-1}\Gamma _{2l\uparrow },  \label{gke}
\end{equation}%
The relations for spin down operators can be obtained
replacing $\uparrow$ by $\downarrow$ and the imaginary unit $i$ by $-i$.

For a finite chain of an even number of sites $L$, the zero-energy end modes 
$a_{1\sigma }$, $b_{L\sigma }$, are also split by a perturbative process of
order $t^{L/2}$ that involves the $\Gamma _{j\sigma }$ and $\Gamma _{j\sigma
}^{\dagger }$ with even $j$, as it can be seen from Eqs.~(\ref{conmut}). For
odd $L$, $H_{t}$ mixes $b_{L\sigma }$ with the subspace of the $\Gamma
_{j\sigma }$ and $\Gamma _{j\sigma }^{\dagger }$ with odd $j$ and $%
b_{L\sigma }$ remains decoupled from $a_{1\sigma }$ at finite $t$.

To calculate the perturbative effective coupling between the end modes it is
easier to map the problem introducing kets associated with the annihilation (%
$a$) and creation ($c$) operators

\begin{equation}
c_{\alpha }\leftrightarrow |\alpha a\rangle \text{, }c_{\alpha }^{\dagger
}\leftrightarrow |\alpha c\rangle ,  \label{map}
\end{equation}%
and introduce the Hamiltonian

\begin{multline}
\tilde{H} = \sum_{\beta \alpha }A_{\beta \alpha }|\beta a\rangle \langle
\alpha a|+B_{\beta \alpha }|\beta c\rangle \langle \alpha a|  - \\
-\overline{A}_{\beta \alpha }|\beta c\rangle \langle \alpha c|+\overline{B}%
_{\beta \alpha }|\beta a\rangle \langle \alpha c|,  \label{htilde}
\end{multline}%
with coefficients defined from the equations

\begin{equation}
\lbrack c_{\alpha },H]=\sum_{\beta }(A_{\beta \alpha }c_{\beta }+B_{\beta
\alpha }c_{\beta }^{\dagger }),  \label{com}
\end{equation}%
so that solving $\tilde{H}|\Gamma _{\nu }a\rangle =E_{\nu }|\Gamma _{\nu
}a\rangle $ is equivalent to solve Eq.~(\ref{conmu0}).

In this new language the effective perturbative mixing of the end modes for
spin up can be written in the form

\begin{equation}
\tilde{H}_{m}=V|b_{L\uparrow }a\rangle \langle a_{1\uparrow }a|+\mathrm{H.c.,%
}  \label{hm}
\end{equation}%
where

\begin{equation}
V=\sum \frac{\langle b_{L\uparrow }a|H_{t}|e_{M}\rangle
\left(\prod\limits_{l=1}^{M-1}\langle e_{l+1}|H_{t}|e_{l}\rangle \right)\langle
e_{1}|H_{t}|a_{1\uparrow }a\rangle }{\prod\limits_{l=1}^{M}(-E_{l})},
\label{Vm}
\end{equation}%
and $|e_{l}\rangle $, $E_{l}$ label the two possible intermediate states at each of the $M=L/2-1$ 
sites with even $j$ and the corresponding energies. 
The sum runs over all possible $2^M$ combinations of intermediate states.
The state $|e_{l}\rangle $ is either $|\Gamma _{2l\uparrow }a\rangle $ with
energy $E_{l}=2$ or $|\Gamma _{2l\downarrow }c\rangle $ with energy $E_{l}=-2
$. Using Eq.~(\ref{conmut}) it is easy to see that the contribution of the
sum when all intermediate state correspond to annihilation operators ($%
|\Gamma _{2l\uparrow }a\rangle $) is $-t(-it/4)^{M}$. In addition, each time 
$|\Gamma _{2l\uparrow }a\rangle $ is replaced by $|\Gamma _{2l\downarrow
}c\rangle $, $\ $a factor $(-i)^{2}$ appears because of the change in two
matrix elements which is compensated by a change of sign in $E_{l}$.
Therefore, the $2^{M}$ possibilities of choosing the intermediate states
lead to the same contribution. Thus

\begin{eqnarray}
V &=&-Ee^{i\theta },  \label{vf} \\
E &=&t(t/2)^{M},  \label{ene} \\
e^{i\theta } &=&(-i)^{M}.  \label{phase}
\end{eqnarray}%
The eigenstate of Eq.~(\ref{hm}) with positive energy $E$ is $%
(|a_{1\uparrow }a\rangle +e^{i\theta }|b_{L\uparrow }a\rangle )/\sqrt{2}$
which corresponds to the annihilation operator

\begin{equation}
\gamma _{\uparrow }=\frac{1}{\sqrt{2}}\left( a_{1\uparrow }+e^{i\theta
}b_{L\uparrow }\right) .  \label{gup}
\end{equation}%
From time reversal symmetry, one has for the spin down

\begin{equation}
\gamma _{\downarrow }=\frac{1}{\sqrt{2}}\left( a_{1\downarrow }+e^{-i\theta
}b_{L\downarrow }\right) .  \label{gdown}
\end{equation}%
These results agree with previous ones obtained for a long chain with $\mu =0$
but otherwise arbitrary parameters using an algebraic approach.\cite%
{entangle} Here, the nature of the coupling between the end modes becomes
more transparent.

\subsection{Effect of a magnetic field at one end}
\label{field}

Since the total spin projection in the direction of the spin orbit
coupling $S_{z}$, is a good quantum number, the effect of a uniform magnetic
field in the $z$ direction is trivial and does not modify the eigenstates,
just changing the energies. Instead, a magnetic field applied to only one end
of the chain leads to non-trivial results. It is easy to generalize 
Eqs.~(\ref{gup}) and (\ref{gdown}) to this case, adding to the Hamiltonian the term 
$-\Delta _{Z}S_\text{right}^{z}$, with $\Delta _{Z}=g\mu _{B}B$ and $%
S_\text{right}^{z}=\sum_{j=L/2}^{L}S_{j}^{z}$ (the total spin projection at the right
half of the chain). 
In practice, only the terms in the sum within a distance to the end of the chain less or of the order of the 
localization length of the zero-energy mode contribute, because for  
other sites the singlet character of the superconductor tends to decrease strongly $|S_{j}^{z}|$.
The expectation value of $S_{j}^{z}$ as a function of lattice site $j$ has been studied 
numerically in Ref.~\onlinecite{entangle}.

The result for the annihilation operators and energies is \cite{entangle}

\begin{eqnarray}
\gamma _{\uparrow } &=&\frac{1}{\sqrt{2}}\left( \alpha a_{1\uparrow }+\beta
e^{i\theta }b_{L\uparrow }\right) ,\text{ }E_{\uparrow }=r-\frac{\Delta _{Z}%
}{4},  \notag \\
\gamma _{\downarrow } &=&\frac{1}{\sqrt{2}}\left( \beta a_{1\downarrow
}+\alpha e^{i\theta }b_{L\downarrow }\right) ,\text{ }E_{\downarrow }=r+%
\frac{\Delta _{Z}}{4},  \label{gb}
\end{eqnarray}%
with
\begin{eqnarray}
r &=&\sqrt{\left( \Delta _{Z}/4\right) ^{2}+E^{2}},  \notag \\
\alpha ^{2} &=&\frac{1}{2}+\frac{\Delta _{Z}}{4r},\;\beta^{2}=1-\alpha ^{2}, \text{ and } \alpha ,\beta >0, \notag \\
E &=&
\begin{cases}
 t^{L/2}/2^{L/2-1} & \text{ if }L\text{ even} \\
 0 & \text{ if }L\text{ odd}  
\end{cases}
\label{albet}
\end{eqnarray}

\subsection{Low-energy many-body eigenstates}
\label{hopm}

Let us discuss first the case without any magnetic field and odd or
infinite $L$, so that the end zero modes are not mixed. In this case,
following a similar reasoning that lead to Eq.~(\ref{e1}) one of the states
that are part of the ground state is, for small $t$

\begin{equation}
|e_1\rangle_{t} =N_{t}(\prod\limits_{k}\Gamma _{k\uparrow }^{e}\Gamma
_{k\downarrow }^{e})(\prod\limits_{k}\Gamma _{k\uparrow }^{o}\Gamma
_{k\downarrow }^{o})|0\rangle ,  \label{e1p}
\end{equation}%
where the annihilation operators are given above [see Eqs.~(\ref{gk}), 
(\ref{gko}), (\ref{gke})]. It is important to note that for small $t$ as we
assume, the energies corresponding to all these operators are positive [Eqs.~(\ref{gk})]. 
For each spin, the operators $\Gamma _{k\sigma }^{o}$ and $%
\Gamma _{k\sigma }^{e}$ are related with the local ones $\Gamma _{j\sigma }$
by a unitary matrix $U_{\sigma }$ with coefficients given by Eqs.~(\ref{gko}) and 
(\ref{gke}) and similarly for spin down. Using this transformation and
Eq.~(\ref{e1}), it is easy to see that

\begin{equation}
|e_1\rangle_{t} =\det (U_{\uparrow })\det (U_{\downarrow })|e_1\rangle .
\label{e1p2}
\end{equation}%
Thus, $|e_1\rangle_{t} $ corresponds to the same physical state as $%
|e_1\rangle .$

In the general case, one of the states that is part of the ground state (non
degenerate for finite even $L)$ has an even number of particles and is given by

\begin{equation}
|g_{e}\rangle =N\gamma _{\uparrow }\gamma _{\downarrow }|e_1\rangle ,
\label{ge}
\end{equation}%
where $N$ is a normalization factor and $\gamma _{\sigma}$ are given by Eqs. (\ref{gup}) and (\ref{gdown}). 

The other three low-energy states and their excitation energies with respect
to the ground state for small enough $t$ and $B$ are

\begin{eqnarray}
|l \uparrow \rangle &=& \gamma _{\uparrow }^{\dagger }|g_{e}\rangle ,\; E_{\uparrow },  \notag \\
|l \downarrow \rangle &=&\gamma _{\downarrow }^{\dagger }|g_{e}\rangle , \; E_{\downarrow },  \notag \\
|l_{e}\rangle  &=&\gamma _{\uparrow }^{\dagger }\gamma _{\downarrow
}^{\dagger }|g_{e}\rangle ,\; 2r.  \label{lex}
\end{eqnarray}%
Note that for odd or infinite $L$, $E=0$, which implies $E_{\uparrow }=0$ and a two-fold 
degeneracy remains in the ground state and in the other two low-energy states.
For a finite chain with odd $L$, this degeneracy is broken if $\mu \neq 0$.\cite{entangle} 
In addition, for the two states with even number of particles, $|g_{e}\rangle$
and $|l_{e}\rangle $, the total spin projection $S_{z}=0$, while 
$S_{z}|l\sigma \rangle =(s_{\sigma }/2)|l\sigma \rangle $. Defining 
$S_\text{left}^{z}=S_{z}-S_\text{right}^{z}$, using previous results,\cite{entangle} and the
vanishing of the trace of $S_{z}$ in the low-energy subspace one obtains

\begin{eqnarray}
\langle g_{e}|S_\text{right}^{z}|g_{e}\rangle  &=&-\langle
g_{e}|S_\text{left}^{z}|g_{e}\rangle   \notag \\
&=&-\langle l_{e}|S_\text{right}^{z}|l_{e}\rangle =\langle
l_{e}|S_\text{left}^{z}|l_{e}\rangle   \notag \\
&=&\frac{\Delta _{Z}}{4\sqrt{\left( \Delta _{Z}\right) ^{2}+16E^{2}}}, 
\notag \\
\langle l \uparrow|S_\text{right}^{z}|l\uparrow \rangle &=&\langle l\uparrow
|S_\text{left}^{z}|l\uparrow \rangle   \notag \\
&=&-\langle l\downarrow |S_\text{right}^{z}|l\downarrow \rangle =\langle l\downarrow
|S_\text{left}^{z}|l\downarrow \rangle   \notag \\
&=&1/4  \label{szend}
\end{eqnarray}

The analytical results valid for small $t$ and $B$ for the excitation energies, Eqs.~(\ref{lex}), and the 
spin projections for the left or right part of the chain, Eqs.~(\ref{szend}), agree very well 
with our numerical results for finite chain. In Table~\ref{tabla} we list some of these numerical results for $L=4$.
The numerical excitation energies presented in the table coincide with the analytical results to the
precision  of the former. For the spin projection, there is a small discrepancy. For example, for
$t=0.04$ and $B=0.002$, the analytical result for $S_\text{right}^{z}$ in the ground state is larger 
by $1.0 \times 10^{-4}$. For the excited  state in the even subspace, the discrepancy is near $3 \times 10^{-4}$. 
The accuracy of the analytical results should increase with the length of the chain, since the effective
mixing of the end states scales as $t^{L/2}$ [see Eqs.~(\ref{hm}), (\ref{vf}), and (\ref{ene})].

\begin{widetext}

\begin{table}[h]
\begin{tabular}{|l|c|c|c|c|c|c|c|}
\hline 
\multicolumn{2}{|c|}{\textbf{Cases}} & \textbf{Energy} & $\left\langle S_{z}\right\rangle $ & $\left\langle S_{1z}\right\rangle $ & $\left\langle S_{2z}\right\rangle $ & $\left\langle S_{3z}\right\rangle $ & $\left\langle S_{4z}\right\rangle $\tabularnewline
\hline 
\hline 
\multirow{4}{*}{$\begin{alignedat}{1}t & =0\\ B & =0 \end{alignedat}$} & 
    \multirow{2}{*}{Even} & -6 & 0 & 0 & 0 & 0 & 0\tabularnewline
\cline{3-8} 
 &  & -6 & 0 & 0 & 0 & 0 & 0\tabularnewline
\cline{2-8} 
 & \multirow{2}{*}{Odd} & -6 & 1/2 & 1/4 & 0 & 0 & 1/4\tabularnewline
\cline{3-8} 
 &  & -6 & -1/2 & -1/4 & 0 & 0 & -1/4\tabularnewline
\hline 
\hline 
\multirow{4}{*}{$\begin{alignedat}{1}t & =0.02\\ B & =0\end{alignedat}$} & 
   \multirow{2}{*}{Even} & -6,00075 & 0 & 0 & 0 & 0 & 0\tabularnewline
\cline{3-8} 
 &  & -6,00035 & 0 & 0 & 0 & 0 & 0\tabularnewline
\cline{2-8} 
 & \multirow{2}{*}{Odd} & -6,00055 & 1/2 & 0,249975 & 0,00002499 & 0,00002499 & 0,249975\tabularnewline
\cline{3-8} 
 &  & -6,00055 & -1/2 & -0,249975 & -0,00002499 & -0,00002499 & -0,249975\tabularnewline
\hline 
\hline 
\multirow{4}{*}{$\begin{alignedat}{1}t & =0\\ B & =2\times10^{-4}\end{alignedat}$} & 
    \multirow{2}{*}{Even} & -6,0001 & 0 & -0,25 & -0,0000125 & 0,0000125 & 0,25\tabularnewline
\cline{3-8} 
 &  & -5,9999 & 0 & 0,25 & -0,0000125 & 0,0000125 & -0,25\tabularnewline
\cline{2-8} 
 & \multirow{2}{*}{Odd} & -6,0001 & -1/2 & -0,25 & -0,0000125 & 0,0000125 & -0,25\tabularnewline
\cline{3-8} 
 &  & -5,9999 & 1/2 & 0,25 & -0,0000125 & 0,0000125 & 0,25\tabularnewline
\hline 
\hline 
\multirow{4}{*}{$\begin{alignedat}{1}t & =0.02\\ B & =2\times10^{-4}\end{alignedat}$} & 
   \multirow{2}{*}{Even} & -6,00077 & 0 & -0,111788 & $-1.31948\times10^{-6}$ & $1.31948\times10^{-6}$ & 0,111788\tabularnewline
\cline{3-8} 
 &  & -6,00033 & 0 & 0,111788 & -0,0000236793 & 0,0000236793 & -0,111788\tabularnewline
\cline{2-8} 
 & \multirow{2}{*}{Odd} & -6,00065 & -1/2 & -0,249975 & -0,0000374894 & -0,0000124906 & -0,249975\tabularnewline
\cline{3-8} 
 &  & -6,00045 & 1/2 & 0,249975 & 0,0000124906 & 0,0000374894 & 0,249975\tabularnewline
\hline
\hline 
\multirow{4}{*}{$\begin{alignedat}{1}t & =0.04\\ B & =0\end{alignedat} $} & 
     \multirow{2}{*}{Even} & -6,003 & 0 & 0 & 0 & 0 & 0\tabularnewline
\cline{3-8} 
 &  & -6,0014 & 0 & 0 & 0 & 0 & 0\tabularnewline
\cline{2-8} 
 & \multirow{2}{*}{Odd} & -6,0022 & -1/2 & -0,2499 & -0,0000998403 & -0,0000998403 & -0,2499\tabularnewline
\cline{3-8} 
 &  & -6,0022 & 1/2 & 0,2499 & 0,0000998403 & 0,0000998403 & 0,2499\tabularnewline
\hline 
\hline 
\multirow{4}{*}{$\begin{alignedat}{1}t & =0\\ B & =0.002 \end{alignedat}$} & 
    \multirow{2}{*}{Even} & -6,001 & 0 & -0,25 & -0,000125 & 0,000125 & 0,25\tabularnewline
\cline{3-8} 
 &  & -5,999 & 0 & 0,25 & -0,000125 & 0,000125 & -0,25\tabularnewline
\cline{2-8} 
 & \multirow{2}{*}{Odd} & -6,001 & -1/2 & -0,25 & -0,000125 & 0,000125 & -0,25\tabularnewline
\cline{3-8} 
 &  & -5,999 & 1/2 & 0,25 & -0,000125 & 0,000125 & 0,25\tabularnewline
\hline 
\hline 
\multirow{4}{*}{$\begin{alignedat}{1}t & =0.04\\ B & =0.002 \end{alignedat}$} & 
     \multirow{2}{*}{Even} & -6,00348 & 0 & -0,195124 & -0,0000468943 & 0,0000468943 & 0,195124\tabularnewline
\cline{3-8} 
 &  & -6,00092 & 0 & 0,195124 & -0,000203056 & 0,000203056 & -0,195124\tabularnewline
\cline{2-8} 
 & \multirow{2}{*}{Odd} & -6,0032 & -1/2 & -0,2499 & -0,000224815 & 0,0000251346 & -0,2499\tabularnewline
\cline{3-8} 
 &  & -6,0012 & 1/2 & 0,2499 & -0,0000251346 & 0,000224815 & 0,2499\tabularnewline
\hline 
\end{tabular}
\caption{\label{tabla} Numerical results for a chain of $L=4$ sites.}
\end{table}

\end{widetext}

\section{Summary and discussion}

\label{sum}

We have found the exact analytical solution, for a particular set of parameters,
of a standard model for a finite chain of a time-reversal-invariant topological superconductor.
This allows us to construct the degenerate ground state
of the system which consists of two states with even fermion parity 
and total spin projection $S_z=0$, and two states with odd fermion parity 
and $S_z= \pm 1/2$. The latter two states have spin projections
$S_z= \pm 1/4$ at the ends. If a magnetic field is applied to one end of the chain, 
the former two states are split in two states having expectation values 
$1/4$ at one end and $-1/4$ at the other. In addition, creating a zero-mode at 
one end or at the other one give related results. This property might
be used for teleportation of Majorana fermions,\cite{tele1,tele2}
or spins,\cite{tele3} since the transport of an electron from one end to the other might be avoided. 

Since the ground state is separated from the excited states by a finite gap,
by continuity these properties remain true for a chain of infinite length and general 
values of the parameters. The coefficients of proportionality in some relations
[like Eq.~(\ref{puntas})] change, but not the proportionality itself.
Specifically the number of low-energy eigenstates (four) and the values 
of the good quantum numbers: total spin projection $S_z$, total fermion parity $P$ 
and the related value of $K^2$ where $K$ is time reversal,  
are discrete numbers which 
should keep their values unmodified under changes of the parameters, unless a phase transition 
takes place. A previous study of the fractional spin projection at the ends
indicates that for general parameters, the localization length of the end modes
increases on approaching the non-topological phase and diverges exactly at the
transition point.\cite{entangle}

Concerning the discussion on the ``time-reversal anomaly,'' in spite of
its use to explain different properties of the system,\cite{review} 
some statements should be revised. 

The main effect of a finite $t$ (or a finite chain of length $L$ in a general case)
is to split the two states of even parity in the ground state by a quantity of order 
$\exp(-L/\lambda_e)$, where $\lambda_e$ is the localization length of the end modes.
This localization length increases with increasing $t$. In addition, the absolute value of the fractional 
expectation values of the spin projection at the ends of the even-parity many-body states 
under the application of a magnetic field, which is 1/4 for $t=0$, decreases for finite $t$.  
This might be experimentally detectable.\cite{entangle}

\section*{Acknowledgments}
We thank Liliana Arrachea for helpful discussions.
We acknowledge support from CONICET, and UBACyT, Argentina. 
We are sponsored by PIP 112-201501-00506 of CONICET and PICT 2017-2726 of
ANPCyT.

\end{document}